\def\useextern{}

\documentclass[runningheads,final]{llncs}
\usepackage{ifluatex}
\usepackage{comment}
\usepackage{graphicx}
\usepackage{float}
\usepackage{booktabs}
\usepackage{colortbl}
\usepackage{enumitem} 

\usepackage{hyperref}
\hypersetup{
  unicode,
  colorlinks = true,
  allcolors = black,
  urlcolor = blue,
  pdfborder = 0 0 0,
}

\usepackage{etoolbox}
\ifdef{\genextern}{%
\usepackage[newfloat,draft=false,chapter,finalizecache]{minted}
}{%
\ifdef{\useextern}{%
\usepackage[newfloat,draft=false,chapter,frozencache]{minted}
}{%
\usepackage[newfloat,draft=false,chapter]{minted}
}}%

\ifdef{\genextern}{%
}{%
\ifdef{\useextern}{%
}{%
}}%
\usepackage{tikz}
\usetikzlibrary{external}

\newmintinline{c}{style=bw}
\newmintinline{text}{style=bw}

\usepackage{multirow}
\usepackage{subcaption}
\usepackage{longtable}

\usepackage[nameinlink,capitalize]{cleveref}
\crefname{sublisting}{Listing}{Listings}
\Crefname{sublisting}{Listing}{Listings}

\usepackage{minibox}

\usetikzlibrary{calc}
\usetikzlibrary{decorations}
\usetikzlibrary{decorations.pathreplacing}
\usetikzlibrary{decorations.text}
\usetikzlibrary{decorations.pathmorphing}
\usetikzlibrary{decorations.markings}
\usetikzlibrary{fit}
\usetikzlibrary{arrows}    
\usetikzlibrary{shapes}
\usetikzlibrary{shadows}
\usetikzlibrary{automata}
\usetikzlibrary{positioning}
\usetikzlibrary{petri}
\usetikzlibrary{chains}
\usetikzlibrary{fadings}
\usetikzlibrary{matrix}
\usetikzlibrary{patterns}
\usetikzlibrary{mindmap}
\usetikzlibrary{graphs}			
\usetikzlibrary{quotes}			
\usetikzlibrary{babel}          
\usetikzlibrary{arrows.meta}
\usetikzlibrary{backgrounds}
\usetikzlibrary{bending}

\ifluatex
\fi

\usepackage{pgfplots}
\pgfplotsset{compat=1.12}

\usepackage{subcaption}

\usepackage[textsize=miniscule,obeyDraft]{todonotes}
\makeatletter
\newcommand\miniscule{\@setfontsize\miniscule{4}{5}}
\makeatother

\newcommand\missing[2][]{\todo[inline,
linecolor=red!50!black,backgroundcolor=red!50!white,bordercolor=red!50!black,#1]{#2}}
\newcommand\done[2][]{}

\tikzset{>={Stealth[round,flex,length=5pt 4.5 0.8]}} 
\tikzset{tight/.style={inner sep=0pt,outer sep=0pt,minimum size=0pt}}

\makeatletter
\newsavebox{\my@resizeenv@TempBox}%
\newcommand*{\my@resizeenv@width}{}%
\newenvironment{resizeenv}[1]{%
\renewcommand*{\my@resizeenv@width}{#1}%
\begin{lrbox}{\my@resizeenv@TempBox}%
}{%
\end{lrbox}%
\resizebox{\my@resizeenv@width}{!}{\usebox{\my@resizeenv@TempBox}}%
}%

\newenvironment{resizepar}{%
\begin{resizeenv}{\textwidth}%
}{%
\end{resizeenv}%
}%

\newsavebox{\my@scale@Lrbox}
\newcommand*{\my@scale@Percentage}{}
\newenvironment*{scaleenv}[1]{%
\renewcommand*{\my@scale@Percentage}{#1}%
\begin{lrbox}{\my@scale@Lrbox}%
}{%
\end{lrbox}%
\scalebox{\my@scale@Percentage}{\usebox{\my@scale@Lrbox}}%
}%

\makeatother

\newcommand{\anchor}[2][]{%
\ifthenelse{\equal{#1}{t}}{%
        \tikz[remember picture,overlay,baseline=(#2 anchor.base)] {\node[outer sep=0pt,inner sep=0pt](#2 anchor){\vphantom{MX}}; \coordinate(#2) at (#2 anchor.north);}%
}{%
        \ifthenelse{\equal{#1}{b}}{%
                        \tikz[remember picture,overlay,baseline=(#2 anchor.base)] {\node[outer sep=0pt,inner sep=0pt](#2 anchor){\vphantom{jp}}; \coordinate(#2) at (#2 anchor.south);}%
        }{%
                \ifthenelse{\equal{#1}{m}}{%
                        \tikz[remember picture,overlay,baseline=(#2 anchor.base)] {\node[outer sep=0pt,inner sep=0pt](#2 anchor){\vphantom{MXjp}}; \coordinate(#2) at (#2 anchor.center);}%
                }{%
                        \tikz[remember picture,overlay,baseline=(#2)] {\coordinate (#2);}%
                }%
        }%
}%
}

\begin{document}
\title{Design and Use of Loop-Transformation Pragmas}

\author{Michael Kruse, Hal Finkel}

\institute{Argonne Leadership Computing Facility,\\
Argonne National Laboratory, Lemont, IL 60439, USA\\
\email{mkruse@anl.gov, hfinkel@anl.gov}}

\maketitle

\begin{abstract}
Adding a pragma directive into the source code is arguably easier than rewriting it, for instance for loop unrolling.
Moreover, if the application is maintained for multiple platforms, their difference in performance characteristics may require different code transformations. 
Code transformation directives allow replacing the directives depending on the platform, i.e. separation of code semantics and its performance optimization.

In this paper, we explore the design space (syntax and semantics) of adding such directive into a future OpenMP specification.
Using a prototype implementation in Clang, we demonstrate the usefulness of such directives on a few benchmarks.

\end{abstract}


\section{Introduction}\label{sct:intro}

In scientific computing, but also in most other kinds of applications, the majority of execution time is spent in loops.
When it comes to improving an application's performance, optimizing the hot loops and their bodies is the most obvious strategy.

While code should be written in a way that is the easiest to understand, it will likely not also be the variant the executes the fastest. 
Platform details such as cache hierarchies, data temporal/spatial locality, prefetching, NUMA, SIMD, SIMT, occupancy, branch prediction, parallelism, work-groups, etc. will have a profound impact on application performance such that restructuring the loop can be necessary.
Since an application rarely runs on just a single platform, one may end up in multiple versions of the same code: One that is written without considering hardware details, and (at least) one for each supported platform, possibly even using different programming models.

OpenMP is intended to be a programming model for many architectures, and ideally allows to share the same code for all of them.
It is comparatively low-effort to replace an OpenMP directive, for instance, using the C/C++ preprocessor and OpenMP 5.0 introduced direct support for this via the \cinline{metadirective}.
Currently, this can only change the parallelization, offloading and vectorization decisions, but not the structure of the code itself.

In our last year's contribution~\cite{iwomp18}, we proposed additional directives in OpenMP for transforming loops, e.g. loop fusion/fission, interchange, tiling, unrolling etc.
In this paper, we discuss choices of syntactic and semantic elements (\cref{sct:design}) for such an addition, give and update on our prototype implementation (\cref{sct:implementation}), and demonstrate how loop transformation can be used in applications and the performance improvements (\cref{sct:evaluation}).

\section{Specification Design Considerations}\label{sct:design}

In this section we explore some of the decisions to make for including loop transformation directives into a potential newer OpenMP standard.
By its nature, this cannot be an exhaustive discussion, but a subjective selection of the most important features that came up in discussion with members of the OpenMP language committee members and others.

The first decision to make is whether to include such directive at all. 
Since the ``MP'' in OpenMP stands for ``\textbf{M}ulti\textbf{P}rocessing'', the original targets of OpenMP were (symmetric) multi-core and -socket platforms, and still today most implementations are based on the pthreads API. 
Multiprocessing obviously does not include sequential loop transformations, but this is not per se a reason to exclude such transformations from OpenMP.

For one, there is a need of supporting functionality: The \cinline{collapse} clause has been added in OpenMP 3.0, although it is not directly related to multiprocessing.
OpenACC~\cite{openacc} also supports a tile-clause.
The \cinline{simd} construct has been added in OpenMP 4.0 to exploit instruction-level parallelism, which also is not included in the term multiprocessing.

Second, the scope of OpenMP has extended relative to its original goal. 
With target offloading also introduced in OpenMP 4.0, it also supports accelerators such as GPGPUs and FPGAs.

There are alternatives to not include code transformations into OpenMP, but have compilers support them in one way or another:
\begin{itemize}[topsep=0pt]
\item 
Continue with the current practice of compiler-specific extensions.\\
Without standardization, these will be incompatible to each other.

\item 
Include into a future version the host languages' specifications (C/C++/Fortran).\\
This would compel OpenMP to add clarifications how its directives interact with the host language's directives. 
However, it is questionable whether e.g. the C++ standard committee will add specifications of pragma-directives. 
Even if all host languages add transformation directives, their semantics are unlikely to match, complicating OpenMP compatibility clarifications.

\item
Create a separate language specification using C/C++/Fortran with OpenMP as its host language.\\
This new language would probably diverge from OpenMP over time as each might add features incompatible to each other.
Comparisons can be drawn from OpenACC, which started as an initiative to add accelerator offloading to OpenMP.
\end{itemize}

For the directives themselves, we distinguish three aspects: Syntax, semantics and the available code transformations.
The syntax describes which token streams are accepted by the compiler and the semantics define their meaning. 
Once these base rules have been defined, it should be straightforward to add transformations consistent with these rules.

\subsection{Syntax}\label{sct:syntax}

In our first proposal~\cite{iwomp18}, we suggested the following syntax:
\begin{minted}{c}
#pragma omp [loop(<loopname(s)>)] <transformation> <clauses...>
\end{minted}
i.e. every transformation is a top-level directive.
The \cinline{loop}-clause before the directive could be used to refer to a loop that is not on the following line or the result of another transformation on the next line.
Since then, the OpenMP 5.0 standard was announced which includes a \textinline{loop}-directive. 
Even though a disambiguation is possible because parentheses follow the clause, but not the directive, overloading the keyword might cause confusion.
Hence, we explore alternatives in this section.


\subsubsection{Loop Directive.}

OpenMP 5.0 introduced the loop construct with the goal to give the compiler more freedom on optimization decisions.
The first OpenMP specification was designed with symmetric multiprocessing in mind, but in the era of heterogeneous computing sensible defaults vary widely.

The idea of the loop-directive was to become the new default worksharing construct, since in most cases, or at least before performance-optimizing an application, the programmer does not care about how the body is executed in parallel, as long as the default choice is reasonable.
In future OpenMP revisions, the loop-construct would gain features of the prescriptive worksharing-construct and preferred when adding new features. 
This maxim also applies to transformation-directives.

\subsubsection{Clauses or (Sub-)Constructs.}

A transformation could be either expressed as a construct (as in~\cite{iwomp18}), or as a clause.
Constructs usually indicate to the compiler to do something, whereas clauses pass options to the construct's doing.
Therefore, a clause requires a construct to be added to.

Currently, OpenMP already uses both syntactic elements for what we might consider loop transformations. 
For instance, \cinline{#pragma omp simd} can be seen as a loop transformation that does vectorization.
On the other side, the collapse clause (valid for multiple constructs such as loop, simd, etc.) is a transformation that occurs before the construct's effect.

When using the loop-construct, the transformation could either be a clause like the collapse-clause, or sub-constructs of the loop clause, similarly to the ``omp'' namespace token before any construct.
However, this would be a new syntactic element in OpenMP in contrast to e.g. \cinline{#pragma omp for simd} is a combined construct, each of them can be used independently. 

The order of any OpenMP clause is irrelevant, but transformations carried out in different orders generally result in different loop nests.
This contradiction can be solved by either make such clauses order-dependent, require the compiler to ignore the order and instead apply an heuristic to determine the best order, or disallow multiple transformations on a single pragma.

\missing{repeated \cinline{pragma omp} boilerplate}

If using the (sub-)construct as the primary syntax, clauses can still be allowed as syntactic sugar where it makes sense and does not cause ambiguity. 
Combined constructs could be allowed as well.

\subsubsection{Loop Chains.}

Bertolacci et. al.~\cite{bertolacci18} proposed a loopchain-construct with a schedule-clause.
The \cinline{loopchain} encloses a loop nest to transform with the \cinline{schedule} clause that defines the transformations to apply on the loop nest, as illustrated in the example below (simplified from~\cite{bertolacci18}).
\begin{minted}{c}
#pragma omplc loopchain schedule(tile(10, parallel, serial))
{
  for (int i = lb ; i <= ub ; i += 1)
    A[i] = (B[i-1] + B[i] + B[i+1]);
  for (int i = lb ; i <= ub ; i += 1)
    A[i] = A[i] * (1.0 / 3.0);
}
\end{minted}
Since the schedule applies the loop nest as a whole, the schedule must also specify an operation on parts that are not transformed.
In the excerpt, the non-transformed part is indicated by the \textinline{serial} operator.
If the loop chain is large with many transformations, the schedule clause can quickly become convoluted.


\subsubsection{Referring to Other Loops.}

Some transformations such as tiling and loop fusion consume more than one loop on the next line and replace them with potentially more than one generated loop, which may be consumed by a follow-up transformation. 
For instance, the result of tiling two nested loops are four loops, and we might want the parallelize the outermost, unroll-and-jam one of the middle loops and vectorize the innermost loop.
Therefore, a syntax is needed to refer to loops that are not directly following the transformation directive.

This can either be done by assigning names to loops and referring to them, or with a path selector from the loop that is annotated. 
Loop names/identifiers have been described in~\cite{iwomp18}, but also used by IBM xlc~\cite{xlcmanual} and XLang~\cite{xlang}.

Path selectors are used for node selection in trees, such as XPath~\cite{xpath} on XML.
In some sense, the collapse clause, taking the number of perfectly nested loops as an argument, is such an selector.
With more complex cases, such as ``the third loop inside the following loop nest of two loops``, maintainability becomes a problem: Adding or removing a loop before between the selector and the selected loop requires updating the selector.

\missing{Non-loop code transformations}

\subsection{Semantics}\label{sct:semantics}

\subsubsection{Prescriptive vs. Descriptive.}

Code transformations are inherently prescriptive: When used, the programmer is already working on performance optimization and cares about the executions order.
The loop-construct is designed to be descriptive and, by default, applies the semantics of \cinline{order(concurrent)}, which allows the compiler to reorder the loop as it fits. 
Then changing the order using a loop transformation directive has no meaning: As the \cinline{order(concurrent)} clause allows an arbitrary permutation/thread-distribution, applying a user-defined permutation will have an undetermined result.
It is also a worksharing-construct, meaning that it is meant to be executed in a \textinline{parallel} context. 
Non-worksharing, simple transformed loops would just run redundantly on every thread in the context.

One solution is to introduce new clauses that disable the default descriptive and worksharing behavior, such as \cinline{order(sequential)} and \cinline{noworksharing}.
To avoid this boilerplate to be repeated with every loop construct, they might be implicit when a loop-transformation is defined.

\subsection{Level Of Prescriptiveness.}

To avoid differences in performance when using different compilers, the specification should define the replacement code of a transformation. 
However, for code that is not performance-sensitive (such as edge cases, fallback code and pro- and epilogue), the compiler might retain some freedom.
Taking the tile-construct as an example, the following decisions are not necessarily performance-relevant:
\begin{itemize}
\item Fallback code for rare cases where the transformation would be invalid, such as address range aliasing of two arrays that would cause a change in semantics.
\item Where and how to execute partial tiles at the logical iteration space border: like a full tile but with additional border conditions or separately after/before all full tiles have been executed.
\item If the iteration counter of the first iteration is not zero, divide tiles using the logical or physical iteration space?
\item Assuming only the code inside a tile is performance-relevant, the outer iteration order over tiles does not need to be defined.
\item If the specification allows tiling of non-perfectly nested loops, there is not obvious way to archive this. 
\end{itemize}
A sensible approach could be to leave these decisions to the compiler, but consider adding clauses that fix this behavior.

OpenMP 5.0 already allows non-perfectly nested loops with the \textinline{collapse}-clause and only requires code between the loops to be executed at most as many times as it would be executed if moved inside the innermost loop, but at least as many times as in the original loops nest.
Executing code more often than in the original code might be an unexpected side-effect of tiling.
In the interest of user-friendless, the specification could disallow non-perfectly loop nests, but add a \textinline{nestify} transformation to make this behavior explicit in the code.

\subsubsection{Transformation Order.}

The order in which multiple transformations are applied on the same loop can be either defined the programmer, the specification, or by the compiler.
When defined by the programmer, the order is derived from the syntax.
Otherwise, any order in the source is ignored and either the OpenMP specification has to specify the rule in which order transformations are applied, or it is implementation-defined such that the compiler can apply heuristics to determine the best ordering.

It might be straight-forward with transformations that consume one loop and replace it with another, but not all orderings are valid with other kinds of transformations. 
For instance, loop interchange requires at least two loops and cannot be applied if the previous transformation only returns a single loop.
If the order is user-defined, the compiler can emit an error.
Otherwise, either the OpenMP has to define which order to use, or the compiler developers.

However, performance optimization engineers will unlikely want to leave such decision up to the compiler or specification.
This is because when using transformations, they will try to get a specific result that is optimal on the target platform and without transformation constructs, would write an alternative code path.
A compiler ``improving'' its heuristic in later versions would also not helpful since it might regress the once-archived performance.

\missing{nested loop transformations before outer loop transformations}

\subsubsection{Compatibility with Legacy Directives.}

Several existing constructs and clauses in OpenMP can be interpreted as a loop transformation:
\begin{itemize}
\item The \textinline{for}, \textinline{loop} and \textinline{distribute}-constructs divide loop iterations between threads or teams.
\item The \textinline{sections}-constructs distributes code regions between threads. 
\item The \textinline{simd} construct vectorizes a loop such that multiple input loop iterations are processed by one iteration of a generated loop, similarly to (partial) unrolling.
\end{itemize}
Using this interpretation, applying other transformations to occur before and after the construct should be possible and make a syntax for new transformations that resemble existing transformations preferable.

Furthermore, existing combined constructs can be redefined as a sequence of transformations, instead of a textual definition.
For instance,
\begin{minted}{c}
#pragma omp for simd schedule(static) simdlen(4)
for (int i = 0; i < n; i+=1)
\end{minted}
could be \emph{defined} as 
\begin{minted}{c}
#pragma omp simd simdlen(4)
#pragma omp for schedule(static)
for (int i = 0; i < n; i+=1)
\end{minted}
Note that this is different from 
\begin{minted}{c}
#pragma omp for schedule(static)
#pragma omp simd simdlen(4)
for (int i = 0; i < n; i+=1)
\end{minted}
which might be more efficient if the number of iterations is not a multiple of the vector width.
Using this transformation extension it is possible to choose between the variants.

\missing{collapse}

\subsubsection{Semantic Safety.}
\done{check\_safety}

Generally, the OpenMP specification requires compilers to apply its directives without regard to whether it is semantically valid to do, i.e. the user guarantees that it is.
This ensures that otherwise conservative compilers still honor the OpenMP directive, but can defer the responsibility to the programmer.

In some scenarios the user might want the compiler to do a validity check.
For instance, the programmer might be unsure themselves or the transformation is added by an autotuner trying out different loop transformations without understanding the code.
For these cases, the directives may support options to instruct the compiler to verify semantic validity.

\begin{table}[t]
\newcommand\original{original}
\newcommand\greenapplied{\textcolor{green!50!black}{transformed}}
\newcommand\greenrtc{\textcolor{green!50!black}{rtc}}
\newcommand\orangeapplied{\textcolor{orange}{transformed}}
\newcommand\redapplied{\textcolor{red}{transformed}}
\begin{tabular}{rcccc}
	\toprule
	               & heuristic                     & default        & \cinline{fallback} & \cinline{force}   \\ 
\cmidrule(l){2-2} \cmidrule(lr){3-3} \cmidrule(l){4-4} \cmidrule(l){5-5}                
	  always valid &  \original \emph{or} \greenapplied & \greenapplied  & \greenapplied    & \greenapplied \\
	valid with rtc &  \original \emph{or} \greenrtc     & \orangeapplied & \greenrtc        & warning \\
	       invalid &  \original                         & \redapplied    & warning          & warning \\
	    impossible &  \original                         & warning        & warning          & warning \\
\bottomrule
\end{tabular}
\centering
\caption{Safety modes for transformation directives. \textcolor{green!50!black}{Green} is for safe transformations, \textcolor{red}{red} may have changed the code's semantics as does \textcolor{orange}{orange} but only in corner cases.}
\label{tab:check_safety}
\end{table}

\Cref{tab:check_safety} shows how safety modes handle different situations for applying a code transformation.
``Always valid'' refers to code to which the transformation can be applied without changing its semantics.
In the case of unrolling this is any loop since unrolling cannot change the code's effect (except execution time).
``Valid with rtc'' refers to code that can be transformed under conditions that can be checked dynamically.
For instance, a transformation may require that two memory regions are not overlapping (alias), which can be checked at runtime if the compiler can deduce which addresse ranges are accessed.
``Invalid'' means that the compiler cannot determine a reasonable runtime condition, i.e. must assume that the transformation will change the code's semantics.
``Impossible'' is code that the compiler can structurally impossible to transform, such as reversing a while-loop\footnote{For general while-loops it is impossible to statically deduce which iteration is the last.}.

Note that these categories may depend on compiler capabilities; e.g. a compiler may have deduced the number of iterations of a while-loop.
For the sake of a standardization, OpenMP should define minimum requirements for compilers to support with everything beyond being a quality-of-implementation.

Without OpenMP, the compiler would heuristically determine whether a transformation is profitable or not.
Hence, it might apply it or not (indicated by ``original'' in \cref{tab:check_safety}), but if it does, it has to ensure that the semantics do not change.

The default behavior of OpenMP directives\footnote{Our previous paper~\cite{iwomp18} suggested to use safe semantics as the default, in conflict to the normal OpenMP behavior} is to always apply even if it the code's semantics changes.
It does not add a runtime check, meaning that the program result can also change in the ``Valid with rtc'' case.
The compiler should emit a warning to the user if the transformation could not be applied at all.

With \cinline{fallback} semantics, the compiler must not emit semantically invalid code, but is allowed to generate fallback code in case a runtime condition fails.
Still, it should warn if the transformation-directive had no effect.
In contrast to the heuristic approach, the compiler skips the profitability check and trusts the directive that the transformation is profitable.

\missing{may combine all fallbacks of all loop nests}

Due to the fallbacks, it is still possible that the non-transformed code is executed without compiler warning and surprise the performance engineer. 
Instead \cinline{force} semantics can be used, which guarantees that either the transformed code is executed, or the compiler emits a warning.
An additional \cinline{required} clause could change the warning to an hard error.

Another idea is a \cinline{hint} clause, which informs the compiler that the transformation is valid (i.e. skips the validity check), but still considers the profitability heuristic, possibly with a bump in favor of applying the transformation instead of the compiler's usual conservativeness.

\subsection{Transformations}\label{sct:transforms}

In addition to the general syntax and semantics, the available transformations have to be defined, including when they are applicable and what the result is.
A convenient approach is to think of transformations as replacements: Remove the code it applies to and insert the result instead. 
Any follow-up transformation can apply on the transformed code as if the replacement was written in the source code.
This should happen internally in the compiler, not textually.

In the remainder of the chapter, we try to define a selected set of transformations.

\subsubsection{Loop Peeling.}

Some loop transformations work best when the loop is a multiple of a constant, such as (partial) unrolling, vectorization and tiling.
If this is not the case, some iterations have to be extracted out of the main loop, which by itself is also a transformation.
Unlike to relying on the implicit peeling, explicitly using a peeling transformation allows more options and naming the resulting prologue- and epilogue-loop to be referenced in follow-up transformations.

We can either the first $k$ iterations into an prologue before the loop or the last $k$ iterations into an epilogue after the loop. 
Peeling the first iterations is always possible, but for peeling the last iterations the number of iterations must be known in advance, which is the case of canonical loops as defined by OpenMP.

The number of iterations to peel can either be specified directly as the number $k$ or indirectly as a goal to archive.
A goal can be:
\begin{enumerate}
\item Make remaining main loop have a multiple of a constant number of iterations; useful for the aforementioned transformations.
\item Make the first access to an array aligned; useful for vectorized loads/stores and accesses that are faster when the compiler knows they are aligned.
\end{enumerate}

Peeling might be necessary spanning multiple loops in a loop nests, since transformations like tiling and unroll-and-jam also apply on multiple nested loops.

\subsubsection{Collapse.}

This combines multiple nested loops into a single logical loop that can be referred to by other transformations.
It should not change the execution order of the inner body.
OpenMP added a clause with similar semantics in version 3.0 and even assigns logical iteration numbers to loop body executions.
A collapse loop-transformation would allow using this functionality independently of other constructs.

\subsubsection{Strip- and Stripe-Mining.}

\begin{figure}[t]
\begin{subfigure}[t]{.5\linewidth}
\begin{scaleenv}{0.9}
\begin{tikzpicture}[x={(0.5,0)},y={(0,-0.5)}]
\foreach \x in {0, ..., 11}
    \foreach \y in {0, ..., 5}
        \node[anchor=center,shape=circle,minimum width=3mm,minimum height=3mm,draw=black!90,very thick](node-\x-\y) at (\x,\y) {};

\begin{pgfonlayer}{background}
\node[fit={(node-6-0) (node-8-5)},tight,draw=blue,very thick,fill=blue!50!white,opacity=0.1] {};
\end{pgfonlayer}
\foreach \x/\xx in {6/7, 7/8}  
    \path (node-\x-0) edge[bend left=85,looseness=2.5,->,blue,opacity=0.3] coordinate[midway] (above-\x) (node-\xx-0); 
\node[above=0mm of above-6,blue,opacity=0.3] {3.};

\begin{pgfonlayer}{background}
\node[fit={(node-3-0) (node-5-5)},tight,draw=blue,very thick,fill=blue!50!white,opacity=0.3] {};
\end{pgfonlayer}
\foreach \x/\xx in {3/4, 4/5}  
    \path (node-\x-0) edge[bend left=85,looseness=2.5,->,blue,opacity=0.5] coordinate[midway] (above-\x) (node-\xx-0); 
\node[above=0mm of above-3,blue,opacity=0.5] {2.};

\begin{pgfonlayer}{background}
\node[fit={(node-0-0) (node-2-5)},tight,draw=blue,very thick,fill=blue!50!white,opacity=1] {};
\end{pgfonlayer}
\foreach \x/\xx in {0/1, 1/2}  
    \path (node-\x-0) edge[bend left=85,looseness=2.5,->,blue,opacity=1] coordinate[midway] (above-\x) (node-\xx-0); 
\node[above=0mm of above-0,blue,opacity=1] {1.};
\end{tikzpicture}
\end{scaleenv}
\caption{Strip-mining}\label{fig:stripmine}
\end{subfigure}
\begin{subfigure}[t]{.45\linewidth}
\begin{scaleenv}{0.9}
\begin{tikzpicture}[x={(0.5,0)},y={(0,-0.5)}]
\foreach \x in {0, ..., 11}
    \foreach \y in {0, ..., 5}
        \node[anchor=center,shape=circle,minimum width=3mm,minimum height=3mm,draw=black!90,very thick](node-\x-\y) at (\x,\y) {};

\begin{pgfonlayer}{background}
\node[fit={(node-2-0) (node-2-5)},tight,draw=blue,very thick,fill=blue!50!white,opacity=0.1] {};
\node[fit={(node-6-0) (node-6-5)},tight,draw=blue,very thick,fill=blue!50!white,opacity=0.1] {};
\node[fit={(node-10-0) (node-10-5)},tight,draw=blue,very thick,fill=blue!50!white,opacity=0.1] {};
\end{pgfonlayer}
\foreach \x/\xx in {2/6, 6/10}  
    \path (node-\x-0) edge[bend left=40,->,blue,opacity=0.2] coordinate[midway] (above-\x) (node-\xx-0); 
\node[above=0mm of above-2,blue,opacity=0.2] {3.};

\begin{pgfonlayer}{background}
\node[fit={(node-1-0) (node-1-5)},tight,draw=blue,very thick,fill=blue!50!white,opacity=0.3] {};
\node[fit={(node-4-0) (node-4-5)},tight,draw=blue,very thick,fill=blue!50!white,opacity=0.3] {};
\node[fit={(node-9-0) (node-9-5)},tight,draw=blue,very thick,fill=blue!50!white,opacity=0.3] {};
\end{pgfonlayer}
\foreach \x/\xx in {1/5, 5/9}  
    \path (node-\x-0) edge[bend left=40,->,blue,opacity=0.5] coordinate[midway] (above-\x) (node-\xx-0); 
\node[above=0mm of above-1,blue,opacity=0.5] {2.};

\begin{pgfonlayer}{background}
\node[fit={(node-0-0) (node-0-5)},tight,draw=blue,very thick,fill=blue!50!white,opacity=1] {};
\node[fit={(node-4-0) (node-4-5)},tight,draw=blue,very thick,fill=blue!50!white,opacity=1] {};
\node[fit={(node-8-0) (node-8-5)},tight,draw=blue,very thick,fill=blue!50!white,opacity=1] {};
\end{pgfonlayer}
\foreach \x/\xx in {0/4, 4/8}  
    \path (node-\x-0) edge[bend left=40,->,blue,opacity=1] coordinate[midway] (above-\x) (node-\xx-0); 
\node[above=0mm of above-0,blue,opacity=1] {1.};
\end{tikzpicture}
\end{scaleenv}
\caption{Stripe-mining}\label{fig:stripemine}
\end{subfigure}
\caption{Mining variants}
\end{figure}
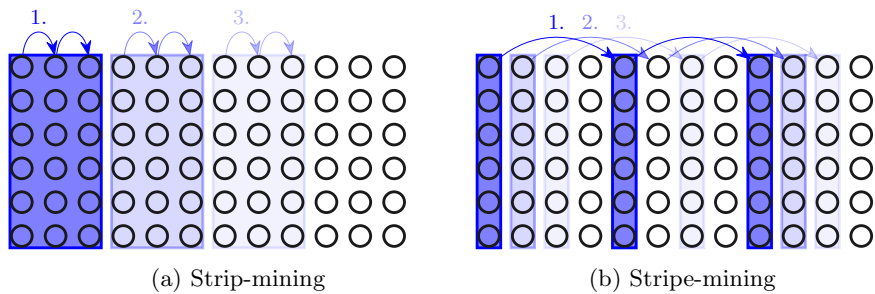

Strip-mining can be seen as one-dimensional tiling.
In contrast to tiling in general, the execution order is not changed, i.e. like unrolling never changes the program's result.
Unlike unrolling, it increases the control-flow complexity and therefore is only intended to be used in combination with other transformations.
For instance, partial unrolling can be implemented by strip-mining followed by a full unroll of the inner loop.
The name is inspired by the term from open-pit mining: The pit is deepened by one strip at a time, as visualized in \cref{fig:stripmine}.

In contrast, stripe-mining does change the execution order: Each inner loop processes a constant number of iterations that are equidistantally distributed over the iteration space.
As shown in \cref{fig:stripemine}, each forms a set of stripes, lending to the transformation's name.

\missing{fusion/fission}
\missing{tiling/strip-mining/stripe-mining/blocking}
\missing{collapse}
\missing{interchange}
\missing{peeling}
\missing{unrolling/interleaving}

\section{Prototype Implementation}\label{sct:implementation}

We created an implementation of some transformation directives in Clang and Polly, which we already described in~\cite{llvmhpc18}.
Because such transformations are not part of OpenMP yet, we use a hybrid of Clang's native syntax for loop transformation extensions and OpenMP construct/clauses syntax. 
The general syntax is:
\begin{minted}{c}
#pragma clang loop(<loopname>) <transforamtion> <clauses...>
\end{minted}

Our code is available on Github\footnote{\url{https://github.com/SOLLVE/clang/tree/pragma} and\\\url{https://github.com/SOLLVE/polly/tree/pragma}} 
Currently, it should be considered as prototype quality and is not intended for use in production. 
For instance, it may crash on syntax errors instead of diagnostic output.

In addition to the transformations mentioned in~\cite{llvmhpc18}, we implemented unrolling, unroll-and-jam, thread-parallelization and peeling for tiled loops.
The parallelization transformation, in contrast to OpenMP's worksharing constructs, can be combined with other transformations.
It should become unnecessary once the interaction between OpenMP's parallelization constructs and loop transformations have been specified.
We unfortunately did not implement loop distribution yet such that it had to be replicated manually for the evaluation.

\missing{OpenMP implementation}
\missing{Attribute implementation}
\missing{Polly implementation}

\section{Evaluation}\label{sct:evaluation}

In this section, we explore how transformation directives can be useful to improve the performance of a selection of kernels.
Please keep in mind that we do not intend to discover new techniques how to improve these kernels over typically hand-optimized kernels in specialized libraries or in literature.
Instead, we want to illustrate how these directives help exploring common optimization techniques.
This is most relevant if no hand-optimized library for the kernel in question is available for a platform.

Unless mentioned otherwise, the execution time was measured on an Intel Core i7 7700HQ (Kaby Lake architecture), 2.8 Ghz with Turbo Boost off and compiled using the \textinline{-ffast-math} switch.
When using parallelism, we use all 8 hardware threads (on 4 physical cores).

\missing{GPU offloading}

\subsection{heat-3d}

\noindent\begin{scaleenv}{0.9}
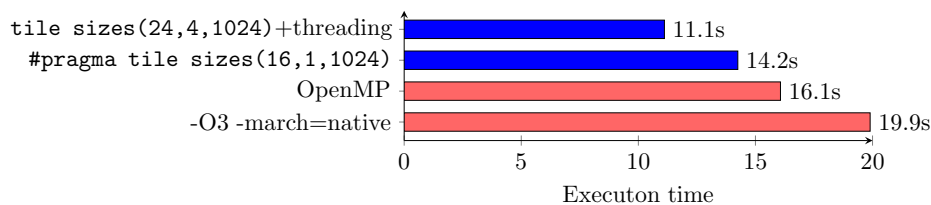
\begin{tikzpicture}
\begin{axis}[
xbar,width=85mm,bar width=1.8ex,bar shift=0pt,height=35mm,
ytick=data,symbolic y coords={base,openmp,pragma_tile,pragma_tile_threaded},yticklabels={{-O3 -march=native},{OpenMP},{\texttt{\#pragma tile sizes(16,1,1024)}},{\texttt{tile sizes(24,4,1024)}+threading}},axis y line=left,
enlarge y limits=0.2,           
xmin=0,xmax=20,xlabel={Executon time},axis x line=bottom,    
nodes near coords={\textcolor{black}{\pgfplotspointmeta}},point meta=explicit symbolic,nodes near coords align={horizontal},nodes near coords style={right} %
]

\addplot[draw=none,fill=none] coordinates {
(0,base)
(0,openmp) 
(0,pragma_tile) 
(0,pragma_tile_threaded)
};

\addplot[draw=black,fill=red!60] coordinates {
(19.897466,base) [19.9s]
(16.068461,openmp) [16.1s]
};

\addplot[draw=black,fill=blue] coordinates {
(14.247319,pragma_tile) [14.2s] 
(11.112794,pragma_tile_threaded) [11.1s] 
};

\end{axis}
\end{tikzpicture}
\end{scaleenv}

The benchmark ``heat-3d'' from Polybench~\cite{polybench421} is 3-dimensional 10-point stencil. 
We are using a volume of $800^3$ and 10 time-steps.
Typical for repeated stencil codes, it alternatingly switches input- and output arrays.
Its 3rd dimension makes it more difficult for the hardware prefetcher.

The baseline can be improved only slightly using OpenMP parallelism (\cinline{#pragma omp parallel for collapse(2)} and \cinline{#pragma omp simd} for the innermost loop).
Tiling improves the performance even more on just a single thread, but can further improved with threading.

The tile sizes were determined using trial-and-error, a task which could also be done by an autotuner.
More advanced time-tiling techniques such as diamond- overlap and tiling and could result in further improvements.

\subsection{syr2k}

Polybench's ``syr2k'' is a rank-2k matrix-matrix update; we are benchmarking matrices of size $4000^2$ and $2000*5200$. 
We run this benchmark on a 2-socket Intel Xeon Gold 6152 CPU (22 cores each, 88 threads in total) with an NVidia Tesla V100-SXM2 GPU.

\noindent
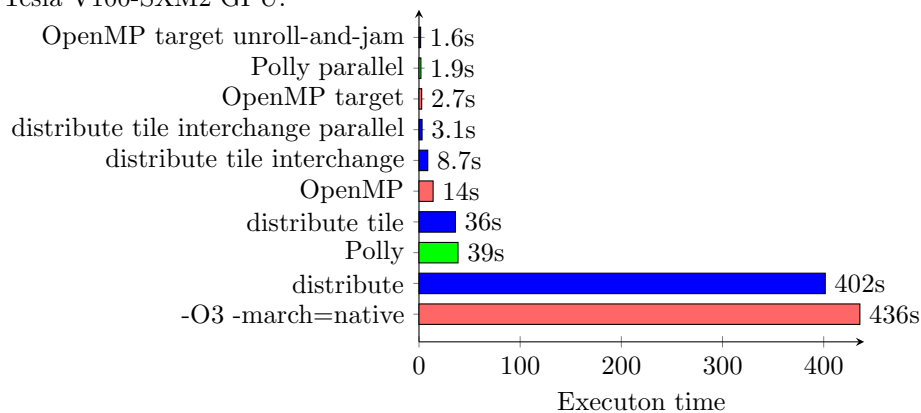
\begin{tikzpicture}
\begin{axis}[
xbar,width=75mm,bar width=1.8ex,bar shift=0pt,height=60mm,
ytick=data,symbolic y coords={
base,
distribute,
polly,
distribute_tile,
openmp,
distribute_tile_interchange,
distribute_tile_interchange_parallel,
gpu,
polly_parallel,
gpu_unrollandjam
},yticklabels={
{-O3 -march=native},
{distribute},
{Polly},
{distribute tile},
{OpenMP},
{distribute tile interchange},
{distribute tile interchange parallel},
{OpenMP target},
{Polly parallel},
{OpenMP target unroll-and-jam}
},axis y line=left,
enlarge y limits=0.1,           
xmin=0,xmax=440,xlabel={Executon time},axis x line=bottom,    
nodes near coords={\textcolor{black}{\pgfplotspointmeta}},point meta=explicit symbolic,nodes near coords align={horizontal},nodes near coords style={right} %
]

\addplot[draw=none,fill=none] coordinates {
(0,base)  
(0,distribute) 
(0,polly)
(0,distribute_tile) 
(0,openmp)  
(0,distribute_tile_interchange)
(0,distribute_tile_interchange_parallel)
(0,gpu)
(0,polly_parallel) 
(0,gpu_unrollandjam) 
};

\addplot[draw=black,fill=red!60] coordinates {
(435.768839,base)  [436s]
(13.961602,openmp) [14s]   
(2.680007,gpu) [2.7s]
};

\addplot[draw=black,fill=blue] coordinates {
(401.548335,distribute) [402s]
(36.022532,distribute_tile) [36s]
(8.677507,distribute_tile_interchange) [8.7s]
(3.056231,distribute_tile_interchange_parallel) [3.1s]
(1.645872,gpu_unrollandjam) [1.6s]
};

\addplot[draw=black,fill=green] coordinates {  
(38.580176,polly) [39s]
(1.911027,polly_parallel) [1.9s]
};

\end{axis}
\end{tikzpicture}

We use the default \cinline{DATASET_EXTRALARGE} for Polybench's ``syr2k''.
In contrast to the stencils, we can gain very high speed-ups.

While loop distribution does not gain a lot by itself, tiling (by 256x96x16) improves the performance by a factor more than 11, followed by a speed-up of another 4x with a loop interchange.
With parallelization on all 44 cores (88 threads), the execution time has improved by a factor of 140 over the original loop.

Interestingly, while single-threaded performance of the Polly-optimized version (using a tile size of 32 in all dimensions and not interchange) is worse, with parallelization it is even better with a speed-up factor of 330.
Evidently, the shared memory bandwidth of the shared caches changes the bottleneck, such that the tile size optimized for single-thread performance is worse. Replication of Polly's optimized loop nest using pragmas replicates the same performance.
We might be able to further improve the performance by searching for a tile size that minimized the traffic higher-level caches.
Using \textinline{#pragma omp parallel for} alone utilizing 88 OpenMP threads yields an improvement of the factor 31.

The performance characteristics changes when offloading to the GPU. With a straightforward \textinline{#pragma omp target teams distribute collapse(2)} of the outer loops and \textinline{#pragma omp parallel for reduction} of the inner loops, the kernel computes in 2.7 seconds, which is slower than the best CPU performance. 
Only with an additional unroll-and-jam did we beat the two CPUs. 
Tiling did not show any improvement.

\subsection{covariance}

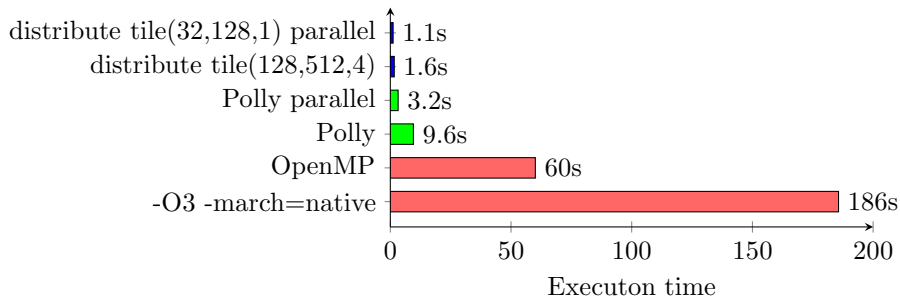
\noindent\begin{tikzpicture}
\begin{axis}[
xbar,width=80mm,bar width=1.8ex,bar shift=0pt,height=45mm,
ytick=data,symbolic y coords={base,openmp,polly,polly_parallel,pragma_tile,pragma_tile_parallel},yticklabels={{-O3 -march=native},{OpenMP},{Polly},{Polly parallel},{distribute tile(128,512,4)},{distribute tile(32,128,1) parallel}},axis y line=left,
enlarge y limits=0.15,           
xmin=0,xmax=200,xlabel={Executon time},axis x line=bottom,    
nodes near coords={\textcolor{black}{\pgfplotspointmeta}},point meta=explicit symbolic,nodes near coords align={horizontal},nodes near coords style={right} %
]

\addplot[draw=none,fill=none] coordinates {
(02,base) 
(0,openmp)
(0,polly)
(0,polly_parallel) 
(0,pragma_tile) 
(0,pragma_tile_parallel) 
};

\addplot[draw=black,fill=red!60] coordinates {
(185.675202,base)  [186s]
(60.075823,openmp) [60s]
};

\addplot[draw=black,fill=blue] coordinates {
(1.630194,pragma_tile)  [1.6s]
(1.157866,pragma_tile_parallel) [1.1s] 
};

\addplot[draw=black,fill=green] coordinates {
(9.564525,polly) [9.6s]
(3.238306,polly_parallel) [3.2s] 
};

\end{axis}
\end{tikzpicture}

The main issue with the covariance benchmarks from Polybench is that the fastest iterator moves the outer data array dimensions leading to strided accesses which cause most of cache lines unused.
If we just transpose the data array (manually), execution time already shrinks to 15 seconds.
The problem can be lessened with tiling.
Unlike the non-tiled version, parallelism improves the execution time only marginally.

Polly's sub-optimal choice of a tile size of 32 for each dimensions also leads to lower performance, for both, the parallel- and single-threaded cases.

\subsection{dgemm}

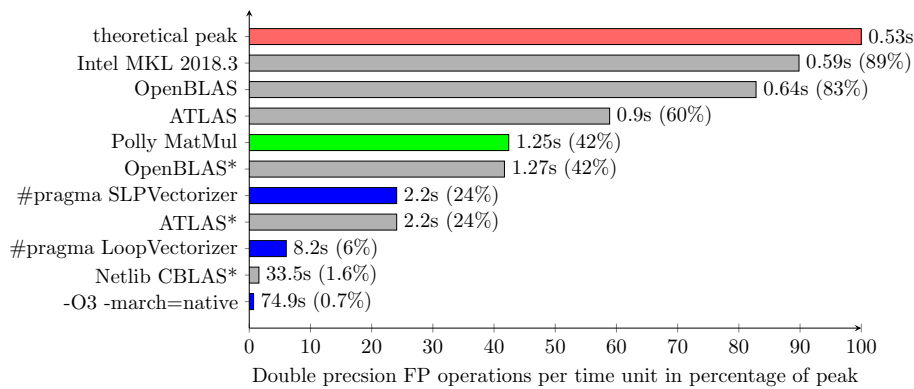
\begin{resizepar}
\begin{tikzpicture}
\begin{axis}[
xbar,width=120mm,height=70mm,bar width=1.8ex,bar shift=0pt,
ytick=data,symbolic y coords={base,cblas,pragma2,atlas,pragma,openblas,polly,atlas_make,openblas_spack,mkl,peak},yticklabels={{-O3 -march=native},{Netlib CBLAS*},{\#pragma LoopVectorizer},{ATLAS*},{\#pragma SLPVectorizer},{OpenBLAS*},{Polly MatMul},{ATLAS},{OpenBLAS},{Intel MKL 2018.3},{theoretical peak}},axis y line=left,enlarge y limits=0.1,           
xmin=0,xmax=100,xlabel={Double precsion FP operations per time unit in percentage of peak},axis x line=bottom,    
nodes near coords={\textcolor{black}{\pgfplotspointmeta}},point meta=explicit symbolic,nodes near coords align={horizontal},nodes near coords style={right} %
]

\addplot[draw=none,fill=none] coordinates {
(0,base) 
(0,cblas)
(0,pragma2)  
(0,atlas) 
(0,pragma)  
(0,openblas)
(0,polly)  
(0,atlas_make)
(0,openblas_spack)
(0,mkl)
(0,peak)   
};

\addplot[draw=black,fill=gray!60] coordinates {
(1.58,cblas)  [33.5s (1.6\%)]
(24.09,atlas)   [2.2s (24\%)] 
(58.8889,atlas_make) [0.9s (60\%)] 
(41.7,openblas)[1.27s (42\%)] 
(82.81,openblas_spack)[0.64s (83\%)] 
(89.83,mkl)     [0.59s (89\%)]
};

\addplot[draw=black,fill=blue] coordinates {
(0.7076,base)   [74.9s (0.7\%)]  
(24.09,pragma)   [2.2s (24\%)] 
(6.05,pragma2) [8.2s (6\%)]
};

\addplot[draw=black,fill=red!60] coordinates {
(100,peak)    [0.53s] 
};

\addplot[draw=black,fill=green] coordinates {
(42.4,polly)    [1.25s (42\%)]
};

\end{axis}
\end{tikzpicture}
\end{resizepar}

In~\cite{iwomp18}, we already optimized Polybench's ``gemm'' kernel, but because of lack of support by LLVM's loop vectorizer, we could only vectorize the innermost loop.
This is sub-optimal because this means that the register dependency is also carried by the innermost loop, restricting the CPU's ability to reorder instructions.

\begin{figure}[h]
\begin{minted}[fontsize=\scriptsize]{c}
#pragma clang loop(i1) pack array(B) \ 
  isl_redirect("{ [c,j,k]   -> [B[x,y] -> PackedB[floord(y,8) mod 256,x mod 256,y mod 8]] }")
#pragma clang loop(j2) pack array(A) \ 
  isl_redirect("{ [c,j,k,l] -> [A[x,y] -> PackedA[floord(x,4) mod 16 ,y mod 256,x mod 4]] }")
#pragma clang loop(i2) unrollingandjam factor(4)
#pragma clang loop(j2) unrollingandjam factor(8)
#pragma clang loop(i1,j1,k1,i2,j2) interchange permutation(j1,k1,i1,j2,i2)
#pragma clang loop(i,j,k) tile sizes(64,2048,256) \
                          floor_ids(i1,j1,k1) tile_ids(i2,j2,k2) peel(rectangular)
  for (int i = 0; i < M; i += 1)
	  for (int j = 0; j < N; j += 1)
		  for (int k = 0; k < K; k += 1)
			  C[i][j] += A[i][k] * B[k][j];
\end{minted}
\caption{Replication of Polly's matrix-multiplication optimization using directives; Libraries marked with (*) were precompiled from the Ubuntu software repository, hence not optimized for the evaluation system}\label{lst:pragma2}
\end{figure}

To avoid this problem, Polly's matrix-multiplication optimization~\cite{gareev18} unroll-and-jams non-inner loops and relies on LLVM's SLP vectorizer to combine the unrolled iterations into vector instructions.
We replicate this behavior in~\cref{lst:pragma2}.
The \cinline{isl_redirect}-clause ensures that the packed arrays' data layout follow the changed access pattern.
For production implementations of the array packing, this should be derived automatically by the compiler.

Unfortunately, the performance is even worse than with the innermost-loop vectorization because, unlike with Polly's output, the SLP vectorizer does not vectorize the jammed loops.
We are working on identifying and fixing the issue in the prototype version.

\subsection{456.hmmer}

\begin{tikzpicture}
\begin{axis}[
xbar,width=70mm,bar width=1.8ex,bar shift=0pt,height=32mm,
ytick=data,symbolic y coords={base,polly,polly_indep},yticklabels={{-O3 -march=native},{Polly \texttt{-polly-stmt-granularity=bb}},{Polly}},axis y line=left,
enlarge y limits=0.4,           
xmin=0,xmax=320,xlabel={Executon time},axis x line=bottom,    
nodes near coords={\textcolor{black}{\pgfplotspointmeta}},point meta=explicit symbolic,nodes near coords align={horizontal},nodes near coords style={right} %
]

\addplot[draw=none,fill=none] coordinates {
(0,base)    
(0,polly)   
(0,polly_indep) 
};

\addplot[draw=black,fill=red!60] coordinates {
(305.69,base)      [306s]
};

\addplot[draw=black,fill=green] coordinates {
(243.31,polly)     [243s]
(166.52,polly_indep) [167s]
};

\end{axis}
\end{tikzpicture}

The most performance-critical code of ``456.hmmer'' from SPEC CPU 2006 is shown in \cref{lst:hmmer}.
Even though it is just one loop, it does 3 independent computations, of which 2 have no loop-carried dependencies.
Separating the sequential computation allows the parallelization and/or vectorization of the two other parts.

\tikzexternaldisable
\begin{figure}[t]
\begin{minted}[escapeinside=??]{c}
for (k = 1; k <= M; k++) {
  ?\anchor[t]{mc1}?mc[k] = mpp[k-1]   + tpmm[k-1];
  if ((sc = ip[k-1]  + tpim[k-1]) > mc[k])  mc[k] = sc;?\anchor[t]{mc2}?
  if ((sc = dpp[k-1] + tpdm[k-1]) > mc[k])  mc[k] = sc;
  if ((sc = xmb  + bp[k])         > mc[k])  mc[k] = sc; 
  mc[k] += ms[k];
  if (mc[k] < -INFTY) mc[k] = -INFTY;?\anchor[b]{mc3}?
  ?\anchor[t]{dc1}?dc[k]?\anchor[b]{dck}? = ?\anchor[t]{dckm1}?dc[k-1]?\anchor[b]{dckm2}? + tpdd[k-1];
  if ((sc = mc[k-1] + tpmd[k-1]) > dc[k]) dc[k] = sc;?\anchor[t]{dc2}?
  if (dc[k] < -INFTY) dc[k] = -INFTY;?\anchor[v]{dc3}?
  if (k < M) {
    ?\anchor[t]{ic1}?ic[k] = mpp[k] + tpmi[k];
    if ((sc = ip[k] + tpii[k]) > ic[k]) ic[k] = sc;?\anchor[t]{ic2}? 
    ic[k] += is[k];
    if (ic[k] < -INFTY) ic[k] = -INFTY;?\anchor[b]{ic3}?
  }
}
\end{minted}
\vspace*{-4ex}
\begin{tikzpicture}[remember picture,overlay]
\node[draw=red,fill=red,inner xsep=1pt,inner ysep=1pt,opacity=0.2,draw opacity=0.5,rounded corners=2pt,fit={(dc1) (dck)}] {};
\node[draw=red,fill=red,inner xsep=1pt,inner ysep=1pt,opacity=0.2,draw opacity=0.5,rounded corners=2pt,fit={(dckm1) (dckm2)}] {};
\node[draw=green,fill=green,inner xsep=1pt,inner ysep=1pt,opacity=0.2,draw opacity=0.5,rounded corners=2pt,fit={(mc1) (mc2) (mc3)},label=right:{Compute \texttt{mc[k]} (\emph{vectorizable})}] {};

\node[draw=red,fill=red,inner xsep=1pt,inner ysep=1pt,opacity=0.2,draw opacity=0.5,rounded corners=2pt,fit={(dc1) (dc2) (dc3)},label=right:{Compute \texttt{dc[k]} (\textbf{not} vectorizable)}] {};
\node[draw=yellow,fill=yellow,inner xsep=1pt,inner ysep=1pt,opacity=0.2,draw opacity=0.5,rounded corners=2pt,fit={(ic1) (ic2) (ic3)},label=right:{Compute \texttt{ic[k]} (\emph{vectorizable})}] {};
\end{tikzpicture}
\caption{456.hmmer hotspot code}\label{lst:hmmer}
\end{figure}
\tikzexternalenable

The figure shows speed-up of the entire 456.hmmer execution (not just the kernel) on an Intel Xeon E5-2667 v3 (Haswell architecture) running at 3.20 GHz.
Earlier versions of Polly only separated one of the computations (using the \textinline{-polly-stmt-granularity=bb} option).
However, the current version separates all 3 computations using its automatic optimizer. 
The same would be possible using a loop distribute directive. 
In contrast to the implicit separation, we could follow-up with additional transformation, such as vectorize one of the loops and parallelize the other.

\section{Conclusion}\label{sct:conclusion}

Loop- --- and more generally: code-transformation directives can be a useful tool to improve a hot code's performance without going too low-level. 
Completely automatic optimizers such as Polly rely on heuristics which are necessarily approximation they do not know the code's dynamic properties (such as number of loop iterations) and have an incomplete performance model of the target machine.
They are also conservative, i.e. rather do nothing than risking performance regressions.

Transformation directives take the burden of profitability analysis off the compile and to the programmer who either knows which transformations are beneficial or can try out multiple approaches, possibly assisted by an autotuner.

We seek to add such transformation directives into a future OpenMP specification, to replace the current compiler-specific pragmas and ensure composability with OpenMP's directives.
We discussed some design choices for syntax and semantics that have to be made with various (dis-)advantages in terms of compatibility, consistency, complexity of implementation and ease of understanding.

\section{Acknowledgments}

This research was supported by the Exascale Computing Project (17-SC-20-SC), a collaborative effort of the U.S. Department of Energy Office of Science and the National Nuclear Security Administration, in particular its subproject on Scaling OpenMP with LLVM for Exascale performance and portability (SOLLVE).

This research used resources of the Argonne Leadership Computing Facility, which is a DOE Office of Science User Facility supported under Contract DE-AC02-06CH11357.


\bibliographystyle{splncs04}
\bibliography{bibliography}

\begin{thebibliography}{1}
\providecommand{\url}[1]{\texttt{#1}}
\providecommand{\urlprefix}{URL }
\providecommand{\doi}[1]{https://doi.org/#1}

\bibitem{bertolacci18}
Bertolacci, I., Strout, M.M., de~Supinski, B.R., Scogland, T.R.W., Davis, E.C.,
  Olschanowsky, C.: {Extending OpenMP to Facilitate Loop Optimization}. In:
  Evolving OpenMP for Evolving Architectures -- 14th International Workshop on
  OpenMP (IWOMP 2018. September 26-28. Barcelona, Spain). Lecture Notes in
  Computer Science, vol. 11128. Springer (2018).
  \doi{10.1007/978-3-319-98521-3\_4}

\bibitem{xlang}
Donadio, S., Brodman, J., Roeder, T., Yotov, K., Barthou, D., Cohen, A.,
  Garzar{\'a}n, M.J., Padua, D., Pingali, K.: {A Language for the Compact
  Representation of Multiple Program Versions}. In: Proceedings of the 18th
  International Workshop on Languages and Compilers for Parallel Computing
  (LCPC'05). pp. 136--151. Springer (2006)

\bibitem{gareev18}
Gareev, R., Grosser, T., Kruse, M.: {High-Performance Generalized Tensor
  Operations: A Compiler-Oriented Approach}. ACM Trans. Archit. Code Optim.
  \textbf{15}(3),  34:1--34:27 (Sep 2018). \doi{10.1145/3235029}

\bibitem{xlcmanual}
IBM: {IBM XL C/C++ for AIX, V16.1 documentation} (2018)

\bibitem{iwomp18}
Kruse, M., Finkel, H.: {A Proposal for Loop-Transformation Pragmas}. In:
  {Evolving OpenMP for Evolving Architectures}. pp. 37--52. Springer
  International Publishing (2018). \doi{10.1007/978-3-319-98521-3\_3}

\bibitem{llvmhpc18}
Kruse, M., Finkel, H.: {User-Directed Loop-Transformations in Clang}. In: 2018
  IEEE/ACM 5th Workshop on the LLVM Compiler Infrastructure in HPC (LLVM-HPC).
  pp. 49--58 (Nov 2018). \doi{10.1109/LLVM-HPC.2018.8639402}

\bibitem{openacc}
OpenACC-Standard.org: {The OpenACC Application Programming Interface Version
  4.0} (Nov 2017)

\bibitem{polybench421}
Pouchet, L.N., Yuki, T.: Polybench 4.2.1 beta,
  \url{https://sourceforge.net/projects/polybench}

\bibitem{xpath}
Spiegel, J., Robie, J., Dyck, M.: {XML Path Language (XPath) 3.1}. {W3C}
  recommendation, W3C (Mar 2017),
  \url{https://www.w3.org/TR/2017/REC-xpath-31-20170321/}

\end{thebibliography}

\end{document}